\documentstyle[editedvolume,epsfig]{crckapb}

\def\epsfsize#1#2{\ifnum#1>\textwidth\textwidth\else#1\fi}

\newcommand{\I}{{\rm i}}

\newcommand{\pD}[2]{\frac{\partial #1}{\partial #2}}

\begin{opening}
\title{Beltrami Equation and Cluster Lensing}
\subtitle{Characteristic Equations \& Applications}
\author{T. Schramm}
\institute{Technical University of Hamburg-Harburg \\
Computer Center -- D 21071 Hamburg \\
and\\
Hamburg Observatory\\
Gojenbergsweg 112 -- D 21029 Hamburg}
\end{opening}

\begin{document}


\begin{abstract}
{Arclets in clusters of galaxies can be used to determine the lens 
mapping and not only to constrain the mass density of the cluster. Multiply 
imaged arclets  are therefore easily identified without further modelling.}
\end{abstract}

\section{The Beltrami equation}
In Schramm \& Kayser (1995) we introduced the complex 
{\bf Beltrami Equation} as an appropriate 
framework for the analysis of arclets in cluster lensing. Corresponding real
formalisms have been developed by Kaiser and Schneider \& Seitz (this volume,
compare also the references in Schramm \& Kayser 1995). 
Here, we show how the solutions of the
Beltrami {\em differential\/} equation can be used to identify multiply imaged
arclets. 
The Beltrami Equation
\begin{equation}
\frac{\partial w}{\partial \bar z}=\mu \frac{\partial w}{\partial z}
\end{equation}
states that a small ellipse in the deflector ($z=x+\I y$) plane given 
by $\mu $ is mapped locally by $w$ onto a circle in the source ($w=u+\I v$) 
plane.
The axial ratio $\epsilon $ of the ellipse and  the direction angle $\varphi$ 
are given by 
$\epsilon =(1-\left| \mu \right| )/(1+\left| \mu \right| )$ and 
$2\varphi=\pi +\arg (\mu )$, respectively.

\section{Known source sizes}
The {\bf Jacobian }is also easily found
\begin{equation}
J=\left| \frac{\partial w}{\partial z}\right| ^2-\left| \frac{\partial w}{%
\partial \bar z}\right| ^2
\end{equation}
Trivially, the mass density $\sigma $ is uniquely determined if the Jacobian
and the Beltrami parameter can be measured at the location of an arclet. 
Since
$
\partial w/\partial z=1-\sigma 
$
we can solve the Beltrami Equation and the Jacobian for the mass density:
\begin{equation}
(1-\sigma )^2=\frac J{1-\left| \mu \right| ^2 }
\end{equation}

\section{Characteristic equations}
Normally we are not so lucky to have the Jacobian but we assume to be able
to measure the $\mu $-field with some accuracy. However, the 
formulation as {\bf differential equation} yields some insights.
For the measurable Beltrami parameter we find for (one-plane) lens mappings
\begin{equation}
\mu =\mu _r+\I\mu_i=\frac{\pD{u}{x}-\pD{v}{y}
+2\I\pD{v}{x}}{\pD{u}{x}+\pD{v}{y}}
\end{equation}
which 
results in two decoupled  linear, homogeneous partial differential equations
\begin{equation}
\mu _i\pD{u}{x}-\left( 1+\mu _r\right) \pD{u}{y} =0\quad, 
\left( 1-\mu _r\right) \pD{v}{x}-\mu _i\pD{v}{y} =0
\end{equation}
The {\bf characteristics} of these equations are given by:
\begin{eqnarray}
x^{\prime }(t)&=&\mu _i,\quad y^{\prime }(t)=-\left( 1+\mu_r\right) 
\Longrightarrow\frac{{\mathrm{d}}y}{{\mathrm{d}}x}=-\frac{1+\mu _r}{\mu _i} \\
x^{\prime }(t)&=&\left( 1-\mu _r\right) ,\quad y^{\prime}(t)=-\mu _i 
\Longrightarrow \frac{{\mathrm{d}}y}{{\mathrm{d}}x}=-\frac{\mu _i}{1-\mu _r}
\end{eqnarray}
where the solutions of these equations are the curves $u,v$=const.
The lens equation is therefore uniquely determined if the  values are 
known at curves (not identical to characteristics). 
Even without this knowledge characteristics are of interest: 
each two (possibly) multiply-{\bf intersecting} characteristics of 
$u$ and $v$ map onto a cross-hair in the source plane so that 
{\bf multiply imaged arclets} can be identified (see example below).

\newpage
\begin{center}
REFERENCES
\end{center}

\noindent
Kaiser, N. 1995, Proc. IAU173, Melbourne, Kluwer Academic Publishers

\noindent
Schneider, P. \& Seitz, S. 1995, Proc. IAU173, Melbourne, Kluwer Academic 
Publishers

\noindent
Schramm, T. \& Kayser, R. 1995, A\&A, {\bf 299}, 1

\begin{figure}[htb]
\centering
\epsfig{file=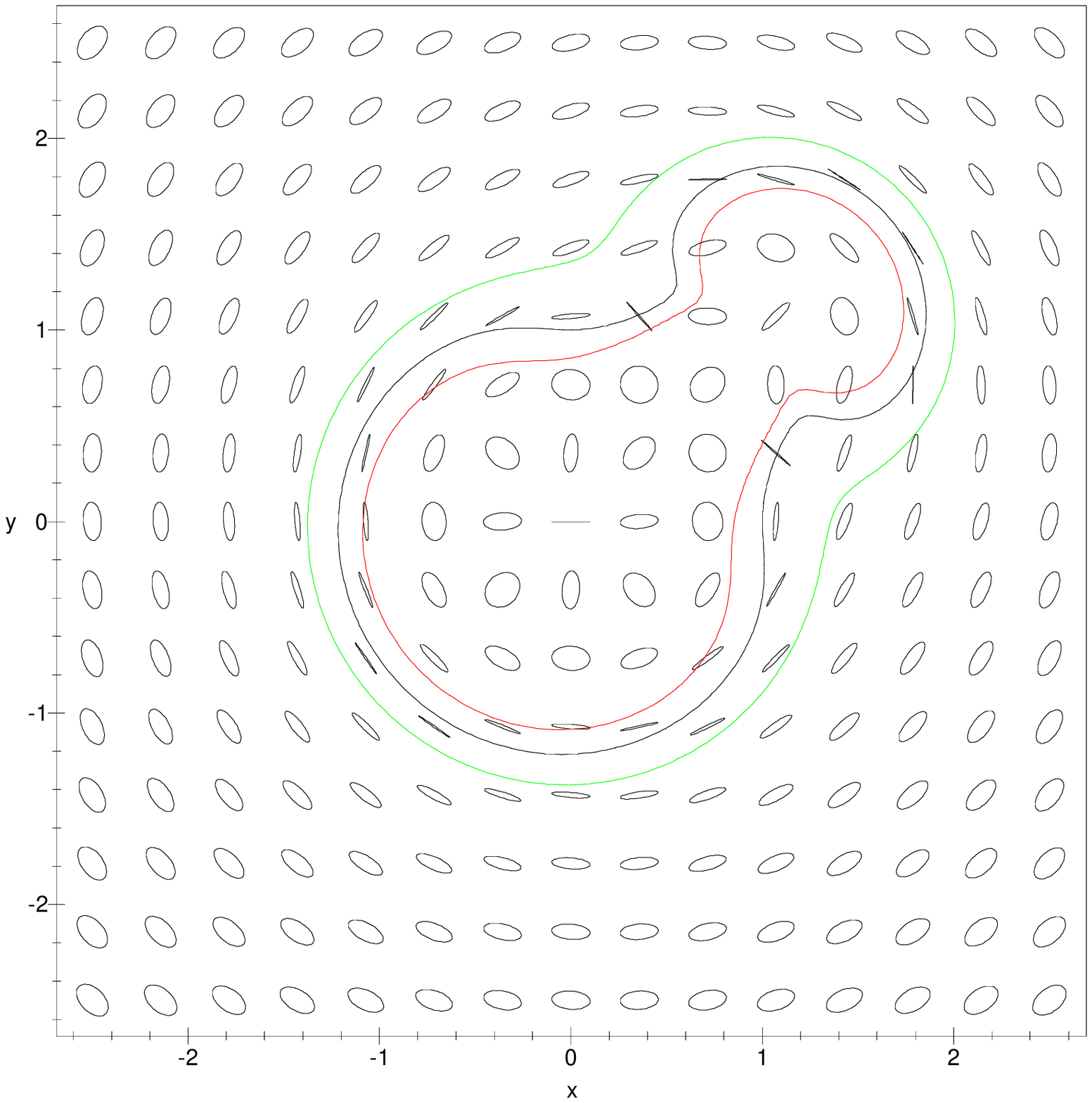,height=12cm,width=12cm,%
bbllx=50pt,bblly=170pt,bburx=520pt,bbury=650pt}
\caption{The ellipse-($\mu$)-field due to a lens composed of two 
different singular isothermal spheres. Additionally three curves 
J=1,0,-1 are plotted. 
}
\end{figure}
\begin{figure}[htb]
\centering
\epsfig{file=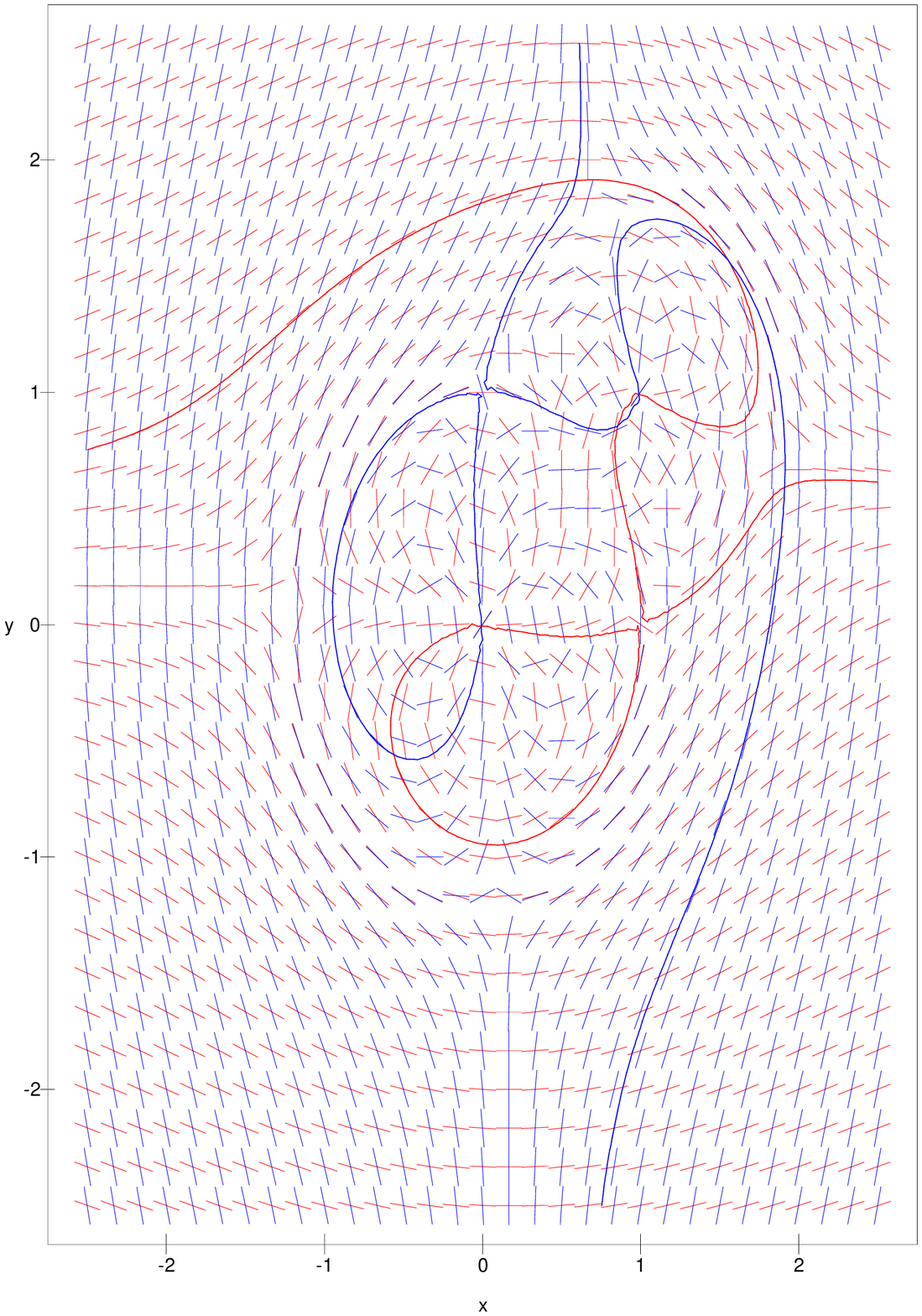,height=12cm,width=12cm,%
bbllx=50pt,bblly=65pt,bburx=520pt,bbury=742pt}
\caption{Right: Overlay of the {\bf direction fields} of the 
differential equations for the curves {$u=$const} and {$v=$const}. 
Additionally two curves {\mbox{$u=c_1$}}, 
{\mbox{$v=c_2$}} are plotted, which are mapped by $w$ 
onto a cross-hair which intersects at 
\mbox{$u+\I v=c_1+\I c_2$} in the $w$-plane.
}
\end{figure}

\end{document}